\documentstyle[prl,aps,epsf,floats]{revtex}
\newcommand{\with}{\,\&\,}
\begin{document}
\draft

\twocolumn[\hsize\textwidth\columnwidth\hsize\csname@twocolumnfalse\endcsname

\title{Bi-defects of Nematic Surfactant Bilayers}
\author{J.-B. Fournier and L. Peliti~\cite{Napoli}}
\address{Laboratoire de Physico-Chimie Th\'eorique, E.\,S.\,P.\,C.\,I.,
10 rue Vauquelin, F-75231 Paris C\'edex 05, France}
\date{\today}
\maketitle
\begin{abstract}
  We consider the effects of the coupling between the orientational
  order of the two monolayers in flat nematic bilayers. We show that the
  presence of a topological defect on one bilayer generates a
  nontrivial orientational texture on both monolayers. Therefore, one
  cannot consider isolated defects on one monolayer, but rather
  associated pairs of defects on either monolayer, which we call
  bi-defects. Bi-defects generally produce walls, such that the
  textures of the two monolayers are identical outside the walls, and
  different in their interior. We suggest some experimental
  conditions in which these structures could be observed.
\end{abstract}
\twocolumn\vskip.5pc]\narrowtext

Nematic liquid crystals are fluid phases possessing a long-range
orientational order~\cite{deGennes}. Ordinary nematics, in the
three-dimensional (3D) space, consist of rodlike molecules orienting
parallel to some unit vector ${\bf n}$, called the ``director''.
Since nematics bear no polar order, ${\bf n}$ and $-{\bf n}$ represent
the same orientational state. Nematics exhibit striking (line or
point) topological defects~\cite{deGennes,Chaikin}. The orientational
order is continuous outside the defect, but exhibits on it a
singularity which cannot be removed by continuous deformations.

Although several almost 2D nematic systems have been investigated,
like thin nematic cells~\cite{Boh?}  and wetting layers~\cite{Braun},
there are few examples of real 2D nematics, e.g., rods suspended on
the surface of aqueous solutions~\cite{Fisch}.  (Actually, 2D systems
can only exhibit {\it quasi\/}-long-range order, but this distinction
is blurred for usual system sizes.)  Very recently, it has been shown
that amphiphilic bilayers made of dimeric surfactants (gemini)
spontaneously form very long tubules of mesoscopic
radius~\cite{Candau}: this conformation can be theoretically explained
by introducing a coupling between the surface curvature and two
independent monolayer nematic orders~\cite{JBF}. A number of
independent arguments support the existence of nematic order in these
membranes~\cite{Candaupriv}.

In this Letter, we investigate the behavior of disclination defects in
such nematic bilayers. For simplicity, we restrict our attention to
planar bilayers, which could be produced by osmotically blowing up the
tubes, or by patch-clamping techniques. We find radically new features
due to the coupling of the nematic order between the two monolayers.
Even if a disinclination is present on only one layer, the coupling
generates a nontrivial texture on the opposite one: this texture
must be considered as a ``defect'' even in the absence of a
singularity. We are thus led to consider pairs of associated defects
on the bilayers, one of which can be virtual (of zero strength).  We
call these structures {\it bi-defects}.

We show that the two interacting nematic monolayers can be mapped on
two independent, ``virtual'', 2D nematic monolayers, one subject
to an external orienting field and the other free. In the
former, defects generate orientational {\it walls}~\cite{Helfrich},
i.e., ribbons where the director turns by $\pi$ on a finite length.
Consequently, bi-defects generally produce walls that reach the
boundary of the sample: the textures of the two monolayers are
identical outside the walls and different in their interior.  The
bi-defect energy is dominated by the walls, and scales therefore
linearly with the sample size (rather than logarithmically).

We denote by ${\bf m}$ and ${\bf n}$ the directors of the upper and
lower monolayer, respectively. Within the one Frank constant
approximation, the nematic free energy of the bilayer can be written
as
\begin{equation}
F=\frac{1}{2}\int\!\! d^2 r\left\{K\left|\nabla {\bf m}\right|^2
+K\left|\nabla {\bf n}\right|^2-\lambda
\left({\bf m}\cdot{\bf n}\right)^2\right\},\label{energy}
\end{equation}
where, e.g., $|\nabla {\bf n}|^2=\partial_i n_j\,\partial_i n_j$
and summation on repeated indices is understood. To be definite,
we suppose $\lambda>0$. This is no restriction,
since there is always the freedom to redefine
${\bf n}$ by a $\pi/2$ rotation, which effectively changes the
sign if the interaction term in Eq.~(\ref{energy}). Let us call
$\theta_+$ (resp.\ $\theta_-$) the polar angle of ${\bf m}$
(resp.\ ${\bf n}$) relative to an arbitrary direction. Setting
$\theta_\pm=\frac{1}{2}(\phi\pm \psi)$,
we obtain (up to an irrelevant additive constant)
$F=\frac{1}{2}(F_0+F_\lambda)$, with
\begin{mathletters}
\begin{eqnarray}
F_0&=&\int d^2r \; \frac{K}{2} \left(\nabla\phi\right)^2;\label{F0}\\
F_\lambda&=&\int d^2r\left\{ \frac{K}{2} \left(\nabla\psi\right)^2
+\lambda\,\sin^2\psi\right\}.\label{Fl}
\end{eqnarray}
Equation (\ref{F0}) describes a free nematic, while
Eq.~(\ref{Fl}) describes a nematic subject to a {\em uniform\/} field directed
along the $\psi=0$ axis~\cite{Aero}. The Euler-Lagrange equation
deriving from (\ref{Fl}) is a sine-Gordon equation:
\end{mathletters}
\begin{equation}
\xi^2\nabla^2(2\psi)=\sin (2\psi),\label{Gordon}
\end{equation}
where $\xi^2=K/(2\lambda)$. The length $\xi$ is the 
analog of the magnetic coherence length of ordinary 
nematics~\cite{deGennes}. The corresponding equation for
$\phi$ is simply $\nabla^2\phi=0$.

A topological defect of strength $p$, located at the origin, is
described in polar coordinates by solutions of the Euler-Lagrange
equations of the form
\begin{equation}
\phi(r,\theta)=p\,\theta +\phi_{\rm c}(r,\theta),\label{defaut}
\end{equation}
(or the analog for $\psi$), where $p$ is a half-integer, and
$\phi_{\rm c}(r,\theta)$ is a continuous function. Indeed, the
director turns by $2p\pi$ in any circuit around the origin. In the
nematic under field, minimization of the energy requires that
$\psi=k\pi$ (where $k$ is an integer) over most of the sample.
Therefore, all nonuniformity is confined within ``soliton'' walls of
thickness $\simeq 5\xi$, crossing which $\psi$ rotates by $\pm
\pi$~\cite{Helfrich,deGennes}.  Thus, a defect of strength $p$
radiates a ``star'' of $2|p|$ walls. Within a region of size $\sim\xi$
around the defect the texture is similar to that without field.

In mean field, the energy of a defect of strength $p$ in a free
nematic is equal to $\pi K p \ln (L/a)$~\cite{deGennes}, where $L$ is
the linear size of the sample and $a$ the radius of a core inside
which the nematic order is destroyed. The interaction energy of two
defects of strength $p_1$ and $p_2$ is given by $-2\pi K p_1 p_2
\ln(d/a)$, where $d$ is the distance between the
defects~\cite{deGennes}. For the nematic under field, the defect
energy is dominated by the energy of the walls, which is equal to
$2K/\xi$ per unit length~\cite{deGennes}.

Since Eq.~(\ref{defaut}) is linear in the defect strength, a bi-defect
$[p,q]$, i.e., the superposition of a defect with a strength $p$ in
the upper monolayer and a strength $q$ in the lower one, is equivalent
to a pair of defects of strength $p+q$ in the free nematic (described
by $\phi$) and of strength $p-q$ in the nematic under field (described
by $\psi$):
\begin{equation}
\left[{p}\atop{q}\right]=\left\{{p+q}\atop{p-q}\right\}.\label{algebra}
\end{equation}
We call $p+q$ the {\it free strength\/} and $p-q$ the {\it field
  strength\/} of the bi-defect.  It follows from our decomposition
that a bi-defect of free strength~$\ell$ and field strength~$m$ obeys
the relations
\begin{mathletters}
\begin{eqnarray}
\theta_\pm\left(\left\{{\ell}\atop{m}\right\}\right)
&=&\frac{1}{2}\big(\theta_0(\ell)\pm\theta_\lambda(m)\big);\\
F\left(\left\{{\ell}\atop{m}\right\}\right)&=&\frac{1}{2}
\big(F_0(\ell)+F_\lambda(m)\big).\label{energie}
\end{eqnarray}
In this equation, $\theta_0(\ell)$ is the texture of a defect of
strength $\ell$ in a free nematic, $\theta_\lambda(m)$ the texture of
a defect of strength $m$ in a nematic under field, and $F_0(\ell)$,
$F_\lambda(m)$ the corresponding energies. In particular,
$\theta_-\left(\left[{p},{0}\right]\right)
=\theta_-\left(\left\{{p},{p}\right\}\right)=\frac{1}{2}
\left(\theta_0(p)-\theta_\lambda(p)\right)$, and therefore there is
a nontrivial texture even in the lower monolayer of a $[p,0]$
bi-defect, where there is no singularity.\end{mathletters}

\begin{figure}
  \centerline{\hspace{0cm}\epsfxsize=8.5cm\epsfbox{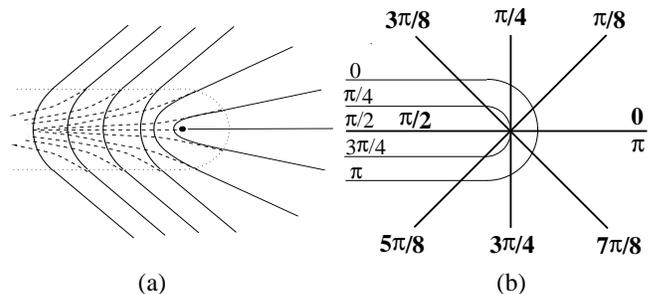}}
  \caption{(a)~Field lines of a $[\frac{1}{2},0]$ bi-defect.
    The textures of the two monolayers coincide outside the wall.
    (b)~Level lines of the corresponding $\{\frac{1}{2},\frac{1}{2}\}$
    bi-defect.}
  \label{demi-0}
\end{figure}

By applying these rules, one can build up the textures corresponding
to different bi-defects. Figure~\ref{demi-0}(a) shows the texture of a
$[\frac{1}{2},0]$ bi-defect. The full (resp.\ dashed) lines are the
field lines~\cite{warning} of the upper (resp.\ lower) monolayer, and
the wall boundary is indicated by the dotted line. The corresponding
$\{\frac{1}{2},\frac{1}{2}\}$ texture is shown in
Fig.~\ref{demi-0}(b): the bold lines are the level lines for the free
nematic, and the thin ones for the field nematic. Figure~\ref{un-0}
shows the analog texture for a $[1,0]$ bi-defect.

\begin{figure}
  \centerline{\hspace{1cm}\epsfxsize=7cm\epsfbox{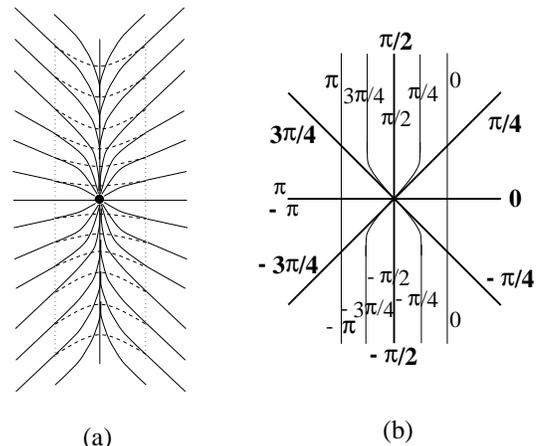}
\hspace{.5cm}}
  \caption{(a)~Field lines of a $[1,0]$ bi-defect. (b)~Level lines
of the corresponding $\{1,1\}$ bi-defect.}
  \label{un-0}
\end{figure}

In crossing a wall, both $\theta_+$ and $\theta_-$ turn by $\pm\pi/2$.
The actual thickness of the walls is $\simeq 5\,\xi$, as one finds by
integrating Eq.~(\ref{Gordon}). Besides, the walls probably have a
persistence length $\xi_{\rm p}$ which is several times their
thickness. Since they are interfaces in two dimensions, they fluctuate
widely: their lateral excursion $\Delta u$ over a length $L$ is given
by
\begin{equation}
\Delta u\simeq \left(\frac{T}{2\pi^2 K}\right)^{1/2}\!(\xi L)^{1/2},
\label{width}
\end{equation}
where $T$ is the temperature, measured in energy units.  They perform
therefore a random walk, but their angular fluctuation $\Delta
\alpha\simeq (T/4 K)^{1/2}(\xi/\xi_{\rm p})^{1/2}$ is small, since we
expect $K$ to be of order a few $T$ in a nematically ordered phase.
The fluctuations of the wall decrease the effective line tension by a
negligible amount.
 
The walls issuing from bi-defects can recombine. Since a defect of
strength $\ell$ under field generates $2|\ell|$ walls, a $[p,q]$
bi-defect generates
\begin{equation}
n=2|p-q|
\end{equation}
walls. Now, if there are two bi-defects, of strengths $[p,q]$ $\&$
$[p',q']$ respectively, the total field strength equals $p-q+p'-q'$
and the number of walls that reach infinity is then $2|p-q+p'-q'|$. If
this number is smaller than $2|p-q|+2|p'-q'|$, some walls must
recombine. This happens if $(p-q)(p'-q')<0$. Therefore, we can assign
an arrow to each wall, pointing {\it outward\/} from the bi-defect if
$(p-q)>0$ and {\it toward\/} it otherwise: walls with matching arrows
can recombine. We show in Fig.~\ref{un-un}(a) the field lines of a
bi-defect pair $[1,0]\with[0,1]$. The two pairs of walls combine,
connecting the two bi-defects, as shown in Fig.~\ref{un-un}(b).

\begin{figure}
  \centerline{\hspace{1cm}\epsfxsize=6.5cm\epsfbox{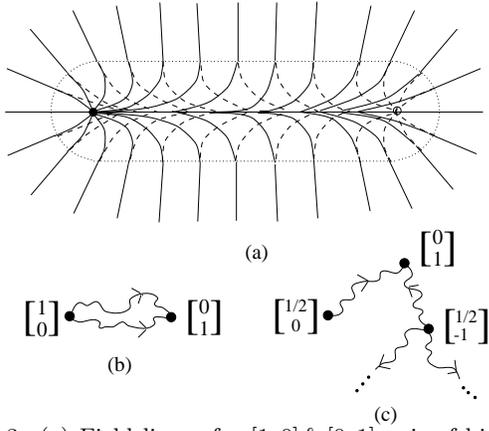}\hspace{1cm}}
  \caption{(a)~Field lines of a $[1,0]\with[0,1]$ pair of bi-defects. 
    (b)~Corresponding scheme of the wall connections. (c)~Possible wall
    connections between three bi-defects.}
  \label{un-un}
\end{figure}

The interaction energy of a $[1,0]\with[0,1]$ bi-defect pair,
equivalent to a $\{1\with1,1\with-1\}$ system, can be estimated using
Eq.~(\ref{energie}). The first contribution, $\frac{1}{2}F_0$, is one
half of the energy of a pair of defects of strength 1 in a free
nematic, i.e., $-\pi K \ln(d/a)$, where $d$ is the distance between
the defects.  The second contribution, $\frac{1}{2}F_\lambda$, is one
half of the energy of the texture under field of a pair of defects of
strength 1 and $-1$.  When $d\gg \xi$ it is dominated by the two walls
which connect the defects, and is therefore $\simeq 4K(d/\xi)$. When
$d\ll \xi$ we can distinguish a region of size $\approx \xi$ where the
texture is similar to that without field, and an exterior region where
$\psi$ is exponentially close to $k\pi$. The corresponding energy
(\ref{Fl}) contains two contributions: the elastic energy $\pi K
\ln(d/a)$ and the potential energy, which is estimated by integrating
$\frac{1}{2}\lambda \sin^2\psi$ for the free texture on a disk of
radius $\approx \xi$. One obtains $\frac{\pi}{4}
K(d/\xi)^2[\ln(\xi/d)+\frac{1}{2}]$. Summing up $\frac{1}{2}F_0$ and
$\frac{1}{2}F_\lambda$ we obtain
\begin{equation}
F_{\rm int}\simeq\cases{\displaystyle\frac{\pi}{4}
K\frac{d^2}{\xi^2}\left[\frac{1}{2}+
\ln\frac{\xi}{d}\right],&for $d\ll\xi$;\cr
\displaystyle4K\frac{d}{\xi},&for $d\gg \xi$.}
\end{equation}
The two bi-defects are therefore attracted by a force which is almost
constant at large separation, and vanishes roughly linearly with $d$
when $d\ll\xi$. Indeed, when the two bi-defects sit on top of
each other, they form a $[1,1]$ bi-defect which optimizes both the
coupling and the elastic energies. 

A $[1,0]\with[0,-1]$ bi-defect pair, equivalent to a
$\{1\with-1,$ $1\with1\}$ system, generates two walls which wander to the
boundary of the sample. The elastic energy, calculated as previously,
is given by
\begin{equation}
F_{\rm int}\simeq\cases{\displaystyle-\frac{\pi}{48}K\frac{d^2}{\xi^2},
&for $d\ll \xi$;\cr
\displaystyle\!\pi K\ln\frac{d}{a}+F_{\rm walls},&for $d\gg \xi$.}
\end{equation}
The energy in the first line is simply the integral of the
$\frac{1}{2}\lambda \sin^2\psi$ term. (The free and field elastic
energies compensate as previously.) There is also a contribution due
to the walls, but it does not depend on $d$.  The first term in the
second line represents the logarithmic attraction of the defects in
the free nematic. $F_{\rm walls}$ is the contribution from the walls
of the nematic under field.  It will depend in general on the way the
walls reach the sample boundary.  Let us consider, e.g., the case in
which the sample is a ribbon of width $2L$, with the two bi-defects in
the middle, each sending a wall to the opposite sides of the ribbon.
Each wall of length $L$ wanders within a rectangular region of width
$\Delta u$ given by Eq.~(\ref{width}). Thus, if $d>\Delta u$, $F_{\rm walls}$
is independent of $d$, whereas, if $d<\Delta u$, there is a
Helfrich-like repulsion between the walls:
\begin{equation}
F_{\rm walls}\approx \frac{T^2}{K}\frac{\xi L}{d^2}.
\end{equation}
Therefore the interaction is repulsive for $d\ll \xi$, and is
otherwise a combination of repulsive and attractive forces, which
identify an equilibrium distance
\begin{equation}
d_{\rm eq}\approx \frac{T}{K}(\xi L)^{1/2}.
\end{equation}

Let us now consider a collection of bi-defects $[p_i,q_i]$
placed in a region of size $R$ inside a sample of size $L\gg R$.
Since the total field strength is given by $\sum_i p_i-\sum_i q_i$,
there are
\begin{equation}
N=2\Big|\sum_ip_i-\sum_iq_i\Big|
\end{equation}
walls going to the boundary. Since the total number of walls issuing from
the defects is $2\sum_i|p_i-q_i|$, there are
\begin{equation}
M=\sum_i\Big|p_i-q_i\Big|-\Big|\sum_ip_i-\sum_iq_i\Big|
\end{equation}
walls linking two bi-defects, that remain confined within $R$.
Therefore the dominant energy, which arises from the walls, scales
as
\begin{equation}
F\approx N\, K\frac{L}{\xi}+M\,K\frac{R}{\xi}.
\end{equation}
In order to minimize its energy, the system will first attempt to
bring $N$ to zero, e.g., by nucleating defects on the
boundary, so as to equalize the total strengths of the defects in
the upper and lower monolayer.  The following step will be to bring together
bi-defects having field strengths of opposite sign, in order
to reduce to $\approx \xi$ the total wall length. The bi-defects
can then recombine.

In unconstrained membranes, there are numerous and subtle effects of
the coupling between in-plane order and curvature (see, e.g.,
\cite{MacLu,LuPro}). Here, in addition, the coupling between the
nematic directors, ${\bf m}$ and ${\bf n}$, and the curvature tensor
${\bf K}$, of the form ${\bf K}:\left({\bf m}\otimes {\bf m}- {\bf
    n}\otimes {\bf n}\right)$~\cite{JBF}, produces interesting but
complicated effects, which are out of the scope of this paper. In
particular, shape fluctuations introduce an effective long-range
coupling between director gradients~\cite{Olmsted}.  On the other
hand, the nematic tends to bend the membrane along its principal
axes~\cite{JBF}.  Therefore the texture around a nematic bi-defect
will deform the membrane, and the membrane shape will react on the
texture in a nontrivial way.

The wall thickness can be estimated by assuming that the $\lambda$
term in Eq.~(\ref{energy}) arises from anisotropic van der Waals
interactions: $\lambda\simeq A_{\rm a}\ell^2/(2\pi d^4)$, where $\ell$
is the linear size of the headgroup, $d$ the membrane thickness, and
$A_{\rm a}$ is the anisotropic Hamaker constant.  Since Hamaker
constants for interactions across a hydrocarbonic medium are of order
$T$~\cite{Israelachvili}, we take $A_{\rm a}\simeq 0.1\,T$. Hence,
with $d\simeq 40\,$\AA\ and $\ell\simeq 10\,$\AA\ we find
$\lambda\simeq 2\,10^{-7}\,$Jm$^{-2}$.  Taking, e.g., $K\simeq 3\,T$,
we obtain $\xi=K^{1/2}/(2\lambda)^{1/2}\simeq 1500\,$\AA.  The wall
thickness, which is of the order of $5\xi$, should be in the $\mu$m
range.

One way to produce flat nematic bilayers would be either to deposit
the membrane on a water-air interface, or to compress a Langmuir
monolayer of gemini until a second layer overlaps the first.
Due to the micrometric thickness of the walls, striking
defect patterns should be directly observable by optical microscopy.

We thank A. Ajdari, P. Olmsted and D. Wu for useful discussions.
LP acknowledges the support of a Chaire Joliot de l'ESPCI.


\begin{references}
\bibitem{warning}There are many equivalent field-line representations
  of the same texture, which are connected by uniform rotations of the
  directors.
\end{references}
\end{document}